\documentclass{iopconfser}
\usepackage{graphicx}
\usepackage{subfig}
\usepackage{xspace}
\usepackage{xcolor}
\definecolor{persianred}{rgb}{0.8, 0.2, 0.2}
\usepackage[numbers,sort&compress]{natbib}
\usepackage{soul}
\usepackage{amsmath}
\usepackage{amssymb}
\usepackage{indentfirst}

\newcommand{\Qibo}{\texttt{Qibo}\xspace}
\newcommand{\Qibojit}{\texttt{Qibojit}\xspace}
\newcommand{\Qibolab}{\texttt{Qibolab}\xspace}
\newcommand{\Qibocal}{\texttt{Qibocal}\xspace}
\newcommand{\Qiboml}{\texttt{Qiboml}\xspace}
\newcommand{\Qibotn}{\texttt{Qibotn}\xspace}
\newcommand{\Qulacs}{\texttt{Qulacs}\xspace}

\newcommand{\quotes}[1]{``#1''}

\begin{document}

\hspace*{\fill} TIF-UNIMI-2024-7

\title{Beyond full statevector simulation with \Qibo}

\author{Andrea Pasquale$^{1,2}$, Andrea Papaluca$^{3}$, Renato M. S. Farias$^{1,4}$, Matteo Robbiati$^{2,5}$, Edoardo Pedicillo$^{1,2}$,
and Stefano Carrazza$^{1,2,5}$}

\affil{$^1$Quantum Research Center, Technology Innovation Institute, Abu Dhabi, UAE}
\affil{$^2$TIF Lab, Dipartimento di Fisica, Università degli Studi di Milano}
\affil{$^3$School of Computing, Australian National University, Canberra, ACT, Australia}
\affil{$^4$Instituto de Física, Universidade Federal do Rio de Janeiro, P.O. Box 68528, Rio de Janeiro, Rio de Janeiro 21941-972, Brazil
}
\affil{$^5$European Organization for Nuclear Research (CERN), Geneva 1211, Switzerland}

\email{andrea.pasquale@unimi.it}

\begin{abstract}
    In this proceedings, we present two new quantum circuit simulation protocols recently added as optional backends to \Qibo, an open-source
    framework for quantum simulation, hardware control and calibration.
    We describe the current status of the framework as for version \texttt{0.2.9}.
    In detail, the two new backends for Clifford and tensor networks simulation are presented and benchmarked against the {\it state-of-the-art}.
\end{abstract}

\section{Introduction}
With the significant achievements reached by quantum technologies~\cite{utility,supremacy, walraff_ec},
the interest towards quantum computing is consistently growing. However, the noise
affecting these early devices is preventing large-scale applications of quantum algorithms.
Waiting for quantum computers to reach a level of reliability that allows for
full-scale execution of quantum computing routines without limitations on the problem size,
it is necessary to continuously improve the simulation tools we have at our disposal.
In particular, the development of classical simulation techniques is crucial to
overcome the intrinsic representation limit in full statevector simulation, where
the memory required to fully represent a qubit system explodes exponentially as
the number of qubits increases.

Several open-source software packages for classical simulation of quantum systems
are available. While some of them, including \texttt{Qiskit}~\cite{qiskit2024} and
\texttt{Cirq}~\cite{cirq_developers_2024_11398048}, \texttt{cuQuantum}~\cite{nvidia},
\texttt{tket}~\cite{tket} and \texttt{Yao}~\cite{yao}, provide general tools for quantum
simulations, other libraries such as \texttt{Pennylane}~\cite{bergholm2022pennylaneautomaticdifferentiationhybrid}
mostly focus on specific applications like quantum machine learning.

\begin{figure}[t!]
  \centering
  \includegraphics[width=\columnwidth]{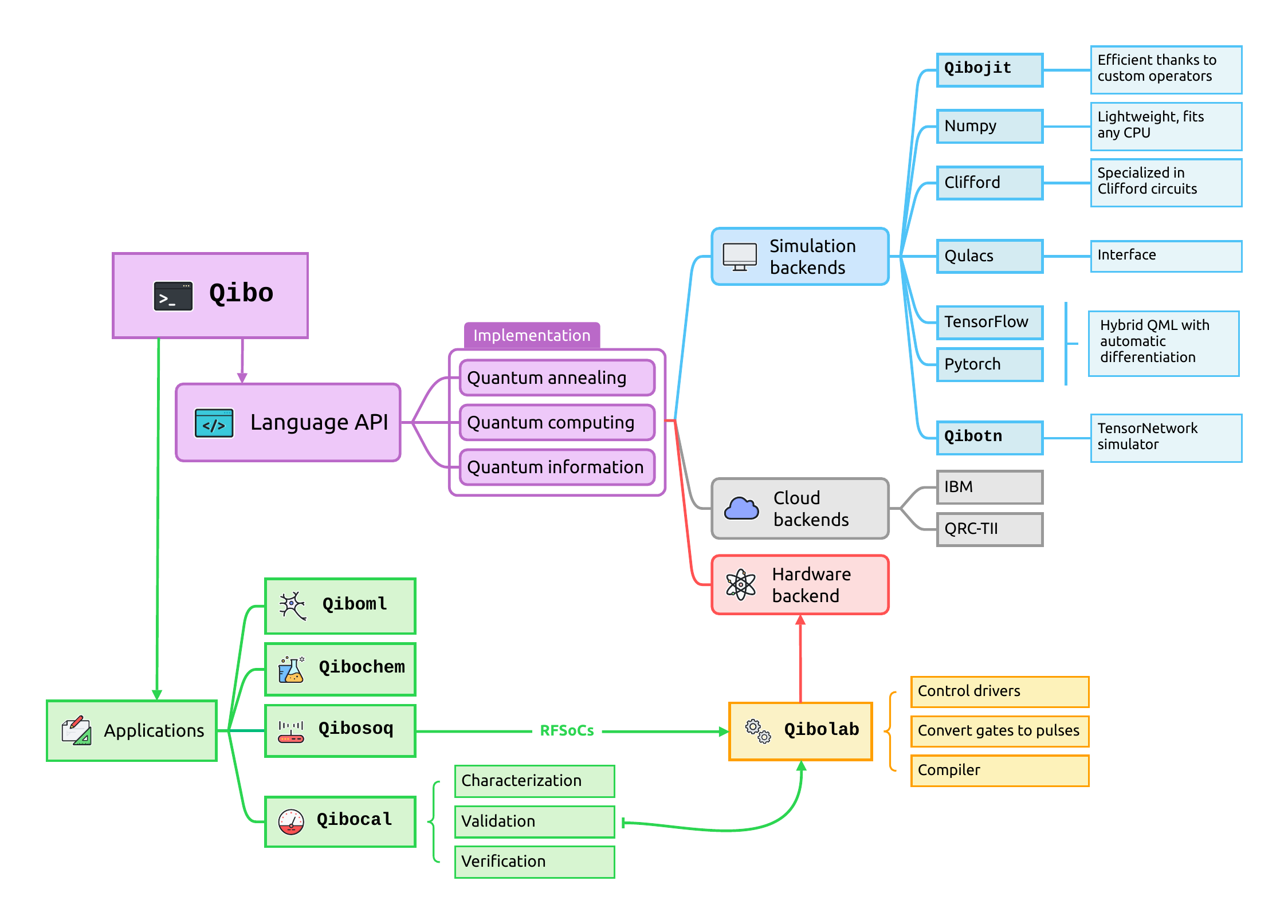}
  \caption{\Qibo framework in version \texttt{0.2.9} \cite{qibo_zenodo}.}
  \label{fig:qibo_ecosystem}
\end{figure}

In this context, for the last 4 years we have been developing \Qibo~\cite{qibo},
an open-source software
framework for quantum computing. At its early stages, \Qibo was mainly dedicated
to full statevector (including density matrix) simulation~\cite{qibo-proceedings},
and managed to achieve performance competitive with state-of-the-art simulators,
thanks, also, to a Just-In-Time compilation approach~\cite{qibojit}.
The current layout of the \Qibo framework, as of version \texttt{0.2.9}, is shown in
Fig.~\ref{fig:qibo_ecosystem}. \Qibo provides a language Application Programming Interface
(API) to deploy quantum algorithms using either a circuit-based or a quantum
annealing-based approach. Moreover, we provide general-purpose tools that are useful in quantum information theory,
including calculation of distances among quantum states and quantum channels.
\Qibo's modular structure enables the deployment of any component of the language
API on different software and hardware platforms, which we refer to
as \emph{backends}.
We are now going to briefly present all the backends available in version \texttt{0.2.9}.
The \Qibo package~\cite{qibo_zenodo} upon installation is equipped with a backend based on
\texttt{Numpy}~\cite{numpy}, which provides adequate performances for simulating circuits with
a relatively low number of qubits $n$ (i.e. $n \leq 20$). We also provide a more efficient
general purpose simulator, called \Qibojit,
which supports hardware acceleration. More specifically, \Qibojit allows multi-threaded CPU execution
using \texttt{Numba}~\cite{numba}, while GPU support is enabled both through \texttt{Cupy}~\cite{cupy}
and through compatibility with NVIDIA's \texttt{CuQuantum}~\cite{nvidia} library.
To improve the performance, \Qibojit defines custom operators that
exploit the sparsity of the matrix representation of several quantum
gates that are often used in quantum computing.
Such improvements are also reflected in the quantum annealing approach when the Trotter decomposition is used.
We also provide backends supporting automatic differentiation specifically designed for quantum machine learning
applications, including one based on \texttt{TensorFlow}~\cite{tensorflow}
primitives, and a recently added one based on the popular machine learning framework \texttt{Pytorch}~\cite{pytorch}.
The possibility of integrating automatic differentiation tools within
such a modular environment can be exploited to develop and test both pure quantum
and hybrid classical-quantum machine learning algorithms~\cite{Bravo_Prieto_2022,
P_rez_Salinas_2021, robbiati2023qaml, robbiati2022quantumanalyticaladamdescent,
robbiati2023realtimeerrormitigationvariational, Cruz_Martinez_2024, particles6010016}.

As a middleware software package, \Qibo also provides a backend for quantum hardware execution.
This backend is called \Qibolab, which defines a dedicated API to perform instrument
control, driver operations, as well as compilation
of \Qibo circuits into customizable native gates sets.
Alongside \Qibolab, we provide a dedicated package to characterize and calibrate
self-hosted quantum devices: \Qibocal~\cite{qibocal_proceedings}.
Besides simulation and hardware backends, there is an ongoing effort to provide cloud access, giving
users the possibility to manage their self-hosted quantum hardware by using \Qibolab's hardware control tools,
or to deploy quantum circuits on well-known quantum cloud providers, including IBM Quantum~\cite{qiskit2024} and IonQ~\cite{IONQCloud}.

This was a brief overview to showcase the wide modularity and diversity that \Qibo provides. As a demonstration of the
ongoing effort to continuously support new platforms and extend \Qibo's compatibility with {\it state-of-the-art} quantum
computing software stacks we recently added a backend that interfaces \Qibo with \Qulacs~\cite{Qulacs}.

Below, we focus on two simulation backends recently added to the \Qibo framework, which are dedicated to tensor network simulation of quantum circuits and fast simulation of Clifford circuits.

\section{Simulation of Clifford circuits}
With full statevector simulation, it is only possible to simulate circuits up to
a limited number of qubits: John Preskill describes this limitation in Ref.~\cite{Preskill_2018}
introducing the approximate yet explanatory term of \textit{50-qubit barrier}.
The main limitations are imposed by  huge memory requirements and very long computation times~\cite{viamontes2003improvinggatelevelsimulationquantum}.
Over the last decades, researchers have studied extensively \emph{Clifford circuits}, a class of circuits that can be simulated in polynomial times~\cite{Aaronson_2004}.
The quantum states generated by Clifford circuits are called \emph{stabilizers states}~\cite{error_correction},
and have many applications in quantum information theory~\cite{Gottesman1997, Gottesman1998, Bravyi2005, Knill2008, Huang2020}.
As current quantum hardware gets closer to the requirements needed for quantum error correction\cite{walraff_ec},
such circuits have started to receive more attention, leading to the development of techniques to classically simulate them.

\begin{figure}[t!]
  \centering
  \subfloat[Simulation of Clifford circuits with an increasing number of qubits.
  For each point we take the average over 100 different randomly generated circuits.
  Circuits were generated following Ref.~\cite{Bravyi_2021}, which guarantees an uniform distribution of the generated $n$-qubit Clifford operators.
  We did not include measurements in this benchmark.]{%
      \includegraphics[width=0.47\textwidth]{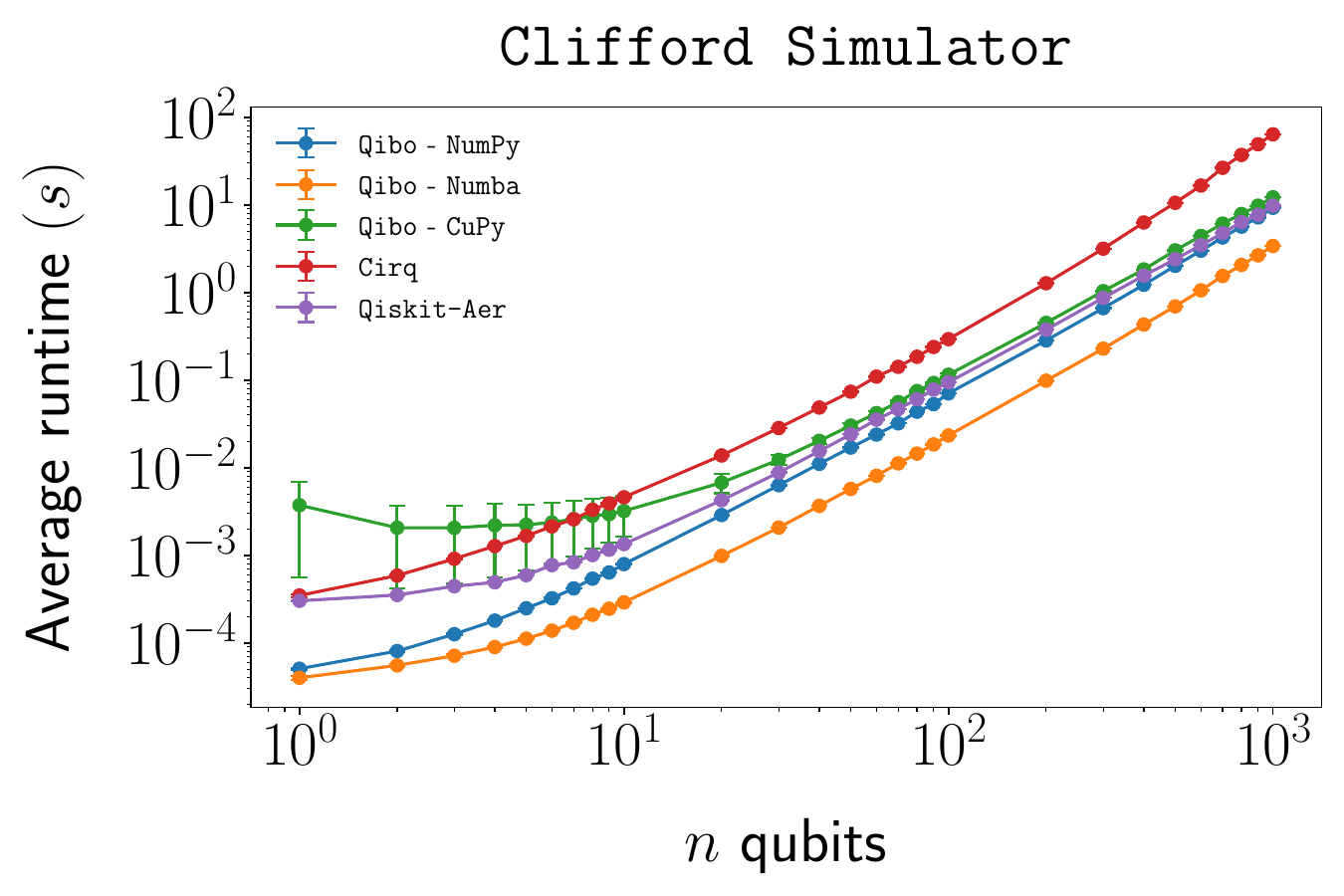}%
      \label{fig:clifford}%
      }%
  \hfill%
  \subfloat[Total simulation time between \Qibojit and \Qibotn for a variational circuit.
  See~\cite{qibojit} for the specific circuit employed.
  The execution is performed on a \texttt{NVIDIA} A100 GPU using NVIDIA and on AMD EPYC 7713.
  ]{%
      \includegraphics[width=0.47\textwidth]{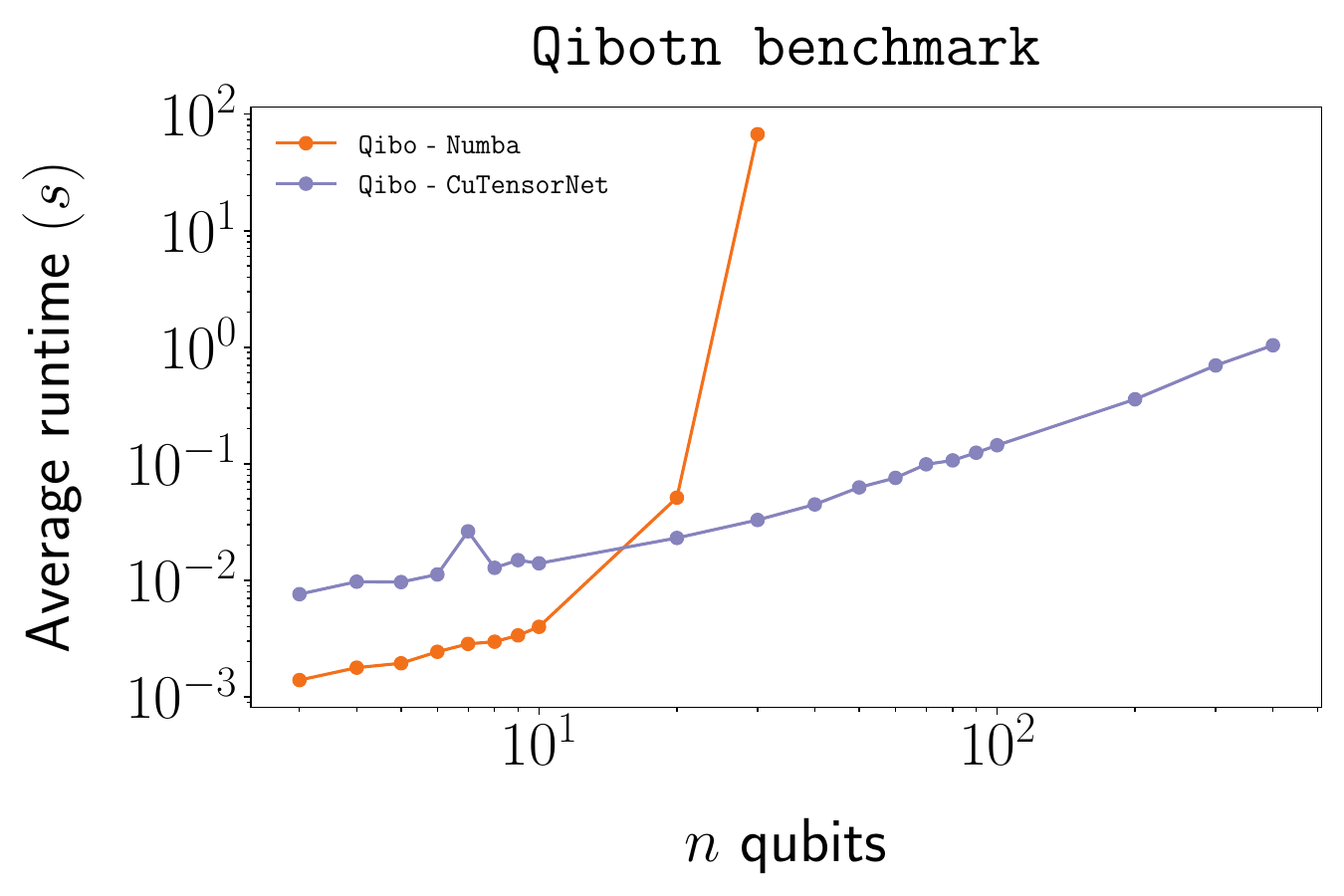}%
      \label{fig:qibotn}%
      }%
  \caption{Benchmarks showing the performances of the Clifford simulator
  and \Qibotn.}
\end{figure}

Based on this, a new backend has been developed in \Qibo which focuses on fast and efficient simulation of Clifford circuits.
The implementation in \Qibo is based on the phase-space formalism introduced in Ref. \cite{Aaronson_2004}.
From the perspective of code design, the implementation makes full use of \Qibo's modularity, allowing for alternative backends to be easily plugged in.
The basic implementation proposed is based on \texttt{Numpy} primitives. This enables the single-threaded simulation of Clifford circuits on CPUs.
By taking advantage of \Qibojit, we also provide implementations based on \texttt{Numba} and \texttt{Cupy}.
The integration with the \texttt{Numba} backend for multi-threaded computations on CPUs is done via custom kernels that are compiled \textit{just-in-time}.
For GPU integration, we make use of custom CUDA C kernels through \texttt{Cupy}.

To evaluate our implementation, we compare our Clifford simulator with Clifford simulators available in \texttt{Qiskit} and \texttt{Cirq}.
The results are shown in Fig.~\ref{fig:clifford}.
We tested all the simulators on the same dataset consisting of $100$ random Clifford circuits for each system syze $n$.
These circuits were sampled uniformly following Ref.~\cite{Bravyi_2021}.
We ran the CPU benchmarks on an AMD EPYC 7773X processor and the GPU benchmarks on a NVIDIA RTX Quadro a6000.
Our \texttt{Numpy} and \texttt{Cupy} backends asymptotically approach performance similar to \texttt{Qiskit},
whereas our \texttt{Numba}-based implementation displays an advantage over the whole range of qubits considered.
For instance, for $n = 1000$, our \texttt{Numba} backend is up to one order of magnitude faster than \texttt{Qiskit}, and almost two order of magnitudes faster than \texttt{Cirq}.

Here, we make a remark about our GPU implementation.
As expected, the overhead of copying the data from the host to the device is dominating the results for a small number of qubits.
However, an improvement due to the huge parallelization capabilities of GPUs is expected to appear at $n \gtrsim 2^{11}$.
Since we benchmarked average runtimes up to $1000$ qubits, this \quotes{crossing} in performance was not observed yet, making further investigation of bigger systems a necessity.

\section{TensorNetwork simulation using \Qibo}

After showing that for specific type of quantum circuits we can reduce the computational time,
we now introduce a second popular approach to simulate large quantum circuits:
classical approximation methods.
A popular method for approximating quantum circuits are tensor networks (TN)~\cite{biamonte2017tensornetworksnutshell},
which represents states or operators as network of smaller tensors reducing both memory and computational requirements.
They are successfully used, for instance, to solve one-dimensional strongly-correlated quantum systems~\cite{Schollw_ck_2011}.
On the other hand, the effectiveness of representing large-scale systems
comes at the cost of introducing truncation errors, which make these techniques less
effective when there is the need to know the quantum state more exhaustively.

Within the \Qibo framework, we have recently developed \Qibotn, a \Qibo subpackage which enables to execute quantum circuits using tensor network like computations,
allowing to support large-scale simulation of quantum circuits in \Qibo.
\Qibotn interfaces \Qibo with state-of-the-art quantum TN simulation libraries such
as \texttt{CuTensorNet}~\cite{nvidia} from NVIDIA and \texttt{quimb}~\cite{quimb}.
Both Matrix Product States and generic TN are supported. \Qibotn is designed to support
High Performance Computing configurations including single node GPUs, as well as multi-node multi-GPU configuration using
Message Passing Interface or the NVIDIA Collective Communication Library (NCCL) from \texttt{NVIDIA}.

To showcase the capabilities of the library we perform comparison between \Qibojit and \Qibotn performances in Fig.\ref{fig:qibotn}.
We execute a variational quantum circuit on a \texttt{NVIDIA} A100 GPU for several number of qubits.
As expected, using full statevector simulation with \Qibojit it possible to run only up to 40,
while using \Qibotn we show how the curve flattens and we observe that we are able to simulate a variational circuit with up to 400 qubits.
Moreover, given the slow rise of the curve we expect to increase the number of qubits with appropriate memory requirements.

\section{Outlook}

In this proceedings, we have described the latest updates available in \Qibo \texttt{0.2.9}.
After a brief overview on all modules currently available in the \Qibo framework, we have put the focus
on two new simulation methodologies: Clifford simulation and Tensor Networks (TN). We have shown that our Clifford simulator
is competitive with {\it state-of-the-art} libraries. We further demonstrated that, despite its early development stage,
our TN implementation is able to simulate efficiently circuits with up to 400 qubits.
Future developments of the \Qibo framework include having a dedicated module to perform QML algorithms,
which we refer to as \Qiboml, and we are looking forward to expand \Qibo to support also quantum chemistry, as well as Quantum optimization problems.
Finally, although \Qibo has been developed in \texttt{Python}, we are looking to separate \Qibo core elements to take advantage of the better performance offered by other languages,
e.g. \texttt{C++} and \texttt{Rust}.

\section{Acknowledgements}
This project is supported by \texttt{TII}’s Quantum Research Center.
The authors thank all \Qibo contributors for helpful discussion and Liwei Yang and Andy Tan Kai Yong
for their support in developing \Qibotn.
M.R. is supported by \texttt{CERN}’s Quantum Technology Initiative (\texttt{QTI}) through the Doctoral Student Program.

\bibliographystyle{unsrtnat}
\bibliography{references}

\begin{thebibliography}{41}
\providecommand{\natexlab}[1]{#1}
\providecommand{\url}[1]{\texttt{#1}}
\expandafter\ifx\csname urlstyle\endcsname\relax
  \providecommand{\doi}[1]{doi: #1}\else
  \providecommand{\doi}{doi: \begingroup \urlstyle{rm}\Url}\fi

\bibitem[Kim et~al.(2023)Kim, Eddins, Anand, Wei, Berg, Rosenblatt, Nayfeh, Wu,
  Zaletel, Temme, and Kandala]{utility}
Youngseok Kim, Andrew Eddins, Sajant Anand, Ken Wei, Ewout Berg, Sami
  Rosenblatt, Hasan Nayfeh, Yantao Wu, Michael Zaletel, Kristan Temme, and
  Abhinav Kandala.
\newblock Evidence for the utility of quantum computing before fault tolerance.
\newblock \emph{Nature}, 618:\penalty0 500--505, 06 2023.
\newblock \doi{10.1038/s41586-023-06096-3}.

\bibitem[Arute et~al.(2019)]{supremacy}
Frank Arute et~al.
\newblock Quantum supremacy using a programmable superconducting processor.
\newblock \emph{Nature}, 574:\penalty0 505–510, 2019.
\newblock \doi{10.1038/s41586-019-1666-5}.
\newblock URL \url{https://doi.org/10.1038/s41586-019-1666-5}.

\bibitem[Krinner et~al.(2022)]{walraff_ec}
Sebastian Krinner et~al.
\newblock Realizing repeated quantum error correction in a distance-three
  surface code.
\newblock \emph{Nature}, 605\penalty0 (7911):\penalty0 669–674, May 2022.
\newblock ISSN 1476-4687.
\newblock \doi{10.1038/s41586-022-04566-8}.
\newblock URL \url{http://dx.doi.org/10.1038/s41586-022-04566-8}.

\bibitem[Javadi-Abhari et~al.(2024)Javadi-Abhari, Treinish, Krsulich, Wood,
  Lishman, Gacon, Martiel, Nation, Bishop, Cross, Johnson, and
  Gambetta]{qiskit2024}
Ali Javadi-Abhari, Matthew Treinish, Kevin Krsulich, Christopher~J. Wood, Jake
  Lishman, Julien Gacon, Simon Martiel, Paul~D. Nation, Lev~S. Bishop,
  Andrew~W. Cross, Blake~R. Johnson, and Jay~M. Gambetta.
\newblock Quantum computing with {Q}iskit, 2024.

\bibitem[Developers(2024)]{cirq_developers_2024_11398048}
Cirq Developers.
\newblock Cirq, May 2024.
\newblock URL \url{https://doi.org/10.5281/zenodo.11398048}.

\bibitem[cuQuantum~development team(2023)]{nvidia}
The cuQuantum~development team.
\newblock Nvidia cuquantum sdk, 2023.
\newblock URL \url{https://github.com/nvidia/cuquantum}.
\newblock BSD-3-Clause License,
  \url{https://github.com/NVIDIA/cuQuantum/blob/main/LICENSE}.

\bibitem[Sivarajah et~al.(2020)Sivarajah, Dilkes, Cowtan, Simmons, Edgington,
  and Duncan]{tket}
Seyon Sivarajah, Silas Dilkes, Alexander Cowtan, Will Simmons, Alec Edgington,
  and Ross Duncan.
\newblock t|ket⟩: a retargetable compiler for nisq devices.
\newblock \emph{Quantum Science and Technology}, 6\penalty0 (1):\penalty0
  014003, November 2020.
\newblock ISSN 2058-9565.
\newblock \doi{10.1088/2058-9565/ab8e92}.
\newblock URL \url{http://dx.doi.org/10.1088/2058-9565/ab8e92}.

\bibitem[Luo et~al.(2020)Luo, Liu, Zhang, and Wang]{yao}
Xiu-Zhe Luo, Jin-Guo Liu, Pan Zhang, and Lei Wang.
\newblock Yao. jl: Extensible, efficient framework for quantum algorithm
  design.
\newblock \emph{Quantum}, 4:\penalty0 341, 2020.

\bibitem[Bergholm
  et~al.(2022)]{bergholm2022pennylaneautomaticdifferentiationhybrid}
Ville Bergholm et~al.
\newblock Pennylane: Automatic differentiation of hybrid quantum-classical
  computations, 2022.
\newblock URL \url{https://arxiv.org/abs/1811.04968}.

\bibitem[Efthymiou et~al.(2021)Efthymiou, Ramos-Calderer, Bravo-Prieto,
  Pérez-Salinas, García-Martín, Garcia-Saez, Latorre, and Carrazza]{qibo}
Stavros Efthymiou, Sergi Ramos-Calderer, Carlos Bravo-Prieto, Adrián
  Pérez-Salinas, Diego García-Martín, Artur Garcia-Saez, José~Ignacio
  Latorre, and Stefano Carrazza.
\newblock \emph{Qibo}: a framework for quantum simulation with hardware
  acceleration.
\newblock \emph{Quantum Science and Technology}, 7\penalty0 (1):\penalty0
  015018, December 2021.
\newblock ISSN 2058-9565.
\newblock \doi{10.1088/2058-9565/ac39f5}.
\newblock URL \url{http://dx.doi.org/10.1088/2058-9565/ac39f5}.

\bibitem[Carrazza et~al.(2023)Carrazza, Efthymiou, Lazzarin, and
  Pasquale]{qibo-proceedings}
S.~Carrazza, S.~Efthymiou, M.~Lazzarin, and A.~Pasquale.
\newblock An open-source modular framework for quantum computing.
\newblock \emph{Journal of Physics: Conference Series}, 2438\penalty0
  (1):\penalty0 012148, February 2023.
\newblock ISSN 1742-6596.
\newblock \doi{10.1088/1742-6596/2438/1/012148}.
\newblock URL \url{http://dx.doi.org/10.1088/1742-6596/2438/1/012148}.

\bibitem[Efthymiou et~al.(2022)Efthymiou, Lazzarin, Pasquale, and
  Carrazza]{qibojit}
Stavros Efthymiou, Marco Lazzarin, Andrea Pasquale, and Stefano Carrazza.
\newblock Quantum simulation with just-in-time compilation.
\newblock \emph{Quantum}, 6:\penalty0 814, September 2022.
\newblock ISSN 2521-327X.
\newblock \doi{10.22331/q-2022-09-22-814}.
\newblock URL \url{http://dx.doi.org/10.22331/q-2022-09-22-814}.

\bibitem[Efthymiou et~al.(2024)Efthymiou, Farias, Carrazza, Papaluca, Robbiati,
  Pedicillo, Pasquale, Bordoni, Sopena, Sam-XiaoyueLi, shangtai, Bravo-Prieto,
  Candido, AdrianPerezSalinas, Vodovozova, Ramos-Calderer, Jun,
  García-Martín, Lazzarin, Khoo, de~Lejarza, Wright, Kong, Zattarin, Puljak,
  Zilli, Paul, Gluza, and rahul]{qibo_zenodo}
Stavros Efthymiou, Renato M.~S. Farias, Stefano Carrazza, Andrea Papaluca,
  Matteo Robbiati, Edoardo Pedicillo, Andrea Pasquale, Simone Bordoni,
  Alejandro Sopena, Sam-XiaoyueLi, shangtai, Carlos Bravo-Prieto, Alessandro
  Candido, AdrianPerezSalinas, Yelyzaveta Vodovozova, Sergi Ramos-Calderer, Wen
  Jun, Diego García-Martín, Marco Lazzarin, Jun~Yong Khoo, Jorge J.~Martínez
  de~Lejarza, Andrew Wright, Jian~Feng Kong, Nicole Zattarin, Ema Puljak, Luca
  Zilli, Paul, Marek Gluza, and rahul.
\newblock qiboteam/qibo: Qibo 0.2.9, June 2024.
\newblock URL \url{https://doi.org/10.5281/zenodo.12577885}.

\bibitem[Harris et~al.(2020)]{numpy}
Charles~R. Harris et~al.
\newblock Array programming with {NumPy}.
\newblock \emph{Nature}, 585\penalty0 (7825):\penalty0 357--362, September
  2020.
\newblock \doi{10.1038/s41586-020-2649-2}.
\newblock URL \url{https://doi.org/10.1038/s41586-020-2649-2}.

\bibitem[Lam et~al.(2015)Lam, Pitrou, and Seibert]{numba}
Siu~Kwan Lam, Antoine Pitrou, and Stanley Seibert.
\newblock Numba: A llvm-based python jit compiler.
\newblock In \emph{Proceedings of the Second Workshop on the LLVM Compiler
  Infrastructure in HPC}, pages 1--6, 2015.

\bibitem[Okuta et~al.(2017)Okuta, Unno, Nishino, Hido, and Loomis]{cupy}
Ryosuke Okuta, Yuya Unno, Daisuke Nishino, Shohei Hido, and Crissman Loomis.
\newblock \emph{CuPy}: A numpy-compatible library for nvidia gpu calculations.
\newblock In \emph{Proceedings of Workshop on Machine Learning Systems
  (LearningSys) in The Thirty-first Annual Conference on Neural Information
  Processing Systems (NIPS)}, 2017.
\newblock URL \url{http://learningsys.org/nips17/assets/papers/paper\_16.pdf}.

\bibitem[Abadi et~al.(2015)]{tensorflow}
Mart\'{i}n Abadi et~al.
\newblock \emph{TensorFlow}: Large-scale machine learning on heterogeneous
  systems, 2015.
\newblock URL \url{https://www.tensorflow.org/}.
\newblock Software available from tensorflow.org.

\bibitem[Paszke et~al.(2019)]{pytorch}
Adam Paszke et~al.
\newblock \emph{PyTorch}: An imperative style, high-performance deep learning
  library, 2019.
\newblock URL \url{https://arxiv.org/abs/1912.01703}.

\bibitem[Bravo-Prieto et~al.(2022)Bravo-Prieto, Baglio, Cè, Francis,
  Grabowska, and Carrazza]{Bravo_Prieto_2022}
Carlos Bravo-Prieto, Julien Baglio, Marco Cè, Anthony Francis, Dorota~M.
  Grabowska, and Stefano Carrazza.
\newblock Style-based quantum generative adversarial networks for monte carlo
  events.
\newblock \emph{Quantum}, 6:\penalty0 777, August 2022.
\newblock ISSN 2521-327X.
\newblock \doi{10.22331/q-2022-08-17-777}.
\newblock URL \url{http://dx.doi.org/10.22331/q-2022-08-17-777}.

\bibitem[Pérez-Salinas et~al.(2021)Pérez-Salinas, Cruz-Martinez, Alhajri, and
  Carrazza]{P_rez_Salinas_2021}
Adrián Pérez-Salinas, Juan Cruz-Martinez, Abdulla~A. Alhajri, and Stefano
  Carrazza.
\newblock Determining the proton content with a quantum computer.
\newblock \emph{Physical Review D}, 103\penalty0 (3), February 2021.
\newblock ISSN 2470-0029.
\newblock \doi{10.1103/physrevd.103.034027}.
\newblock URL \url{http://dx.doi.org/10.1103/PhysRevD.103.034027}.

\bibitem[Robbiati et~al.(2023{\natexlab{a}})Robbiati, Cruz-Martinez, and
  Carrazza]{robbiati2023qaml}
Matteo Robbiati, Juan~M. Cruz-Martinez, and Stefano Carrazza.
\newblock Determining probability density functions with adiabatic quantum
  computing, 2023{\natexlab{a}}.
\newblock URL \url{https://arxiv.org/abs/2303.11346}.

\bibitem[Robbiati et~al.(2022)Robbiati, Efthymiou, Pasquale, and
  Carrazza]{robbiati2022quantumanalyticaladamdescent}
Matteo Robbiati, Stavros Efthymiou, Andrea Pasquale, and Stefano Carrazza.
\newblock A quantum analytical adam descent through parameter shift rule using
  qibo, 2022.
\newblock URL \url{https://arxiv.org/abs/2210.10787}.

\bibitem[Robbiati et~al.(2023{\natexlab{b}})Robbiati, Sopena, Papaluca, and
  Carrazza]{robbiati2023realtimeerrormitigationvariational}
Matteo Robbiati, Alejandro Sopena, Andrea Papaluca, and Stefano Carrazza.
\newblock Real-time error mitigation for variational optimization on quantum
  hardware, 2023{\natexlab{b}}.
\newblock URL \url{https://arxiv.org/abs/2311.05680}.

\bibitem[Cruz-Martinez et~al.(2024)Cruz-Martinez, Robbiati, and
  Carrazza]{Cruz_Martinez_2024}
Juan~M Cruz-Martinez, Matteo Robbiati, and Stefano Carrazza.
\newblock Multi-variable integration with a variational quantum circuit.
\newblock \emph{Quantum Science and Technology}, 9\penalty0 (3):\penalty0
  035053, June 2024.
\newblock ISSN 2058-9565.
\newblock \doi{10.1088/2058-9565/ad5866}.
\newblock URL \url{http://dx.doi.org/10.1088/2058-9565/ad5866}.

\bibitem[Bordoni et~al.(2023)Bordoni, Stanev, Santantonio, and
  Giagu]{particles6010016}
Simone Bordoni, Denis Stanev, Tommaso Santantonio, and Stefano Giagu.
\newblock Long-lived particles anomaly detection with parametrized quantum
  circuits.
\newblock \emph{Particles}, 6\penalty0 (1):\penalty0 297--311, 2023.
\newblock ISSN 2571-712X.
\newblock \doi{10.3390/particles6010016}.
\newblock URL \url{https://www.mdpi.com/2571-712X/6/1/16}.

\bibitem[Pasquale et~al.(2024)Pasquale, Efthymiou, Ramos-Calderer, Wilkens,
  Roth, and Carrazza]{qibocal_proceedings}
Andrea Pasquale, Stavros Efthymiou, Sergi Ramos-Calderer, Jadwiga Wilkens, Ingo
  Roth, and Stefano Carrazza.
\newblock Towards an open-source framework to perform quantum calibration and
  characterization, 2024.
\newblock URL \url{https://arxiv.org/abs/2303.10397}.

\bibitem[{IonQ Inc.}(2024)]{IONQCloud}
{IonQ Inc.}
\newblock {IonQ} {Q}uantum {C}loud, 2024.
\newblock URL \url{https://ionq.com/quantum-cloud}.

\bibitem[Suzuki et~al.(2021)Suzuki, Kawase, Masumura, Hiraga, Nakadai, Chen,
  Nakanishi, Mitarai, Imai, Tamiya, Yamamoto, Yan, Kawakubo, Nakagawa, Ibe,
  Zhang, Yamashita, Yoshimura, Hayashi, and Fujii]{Qulacs}
Yasunari Suzuki, Yoshiaki Kawase, Yuya Masumura, Yuria Hiraga, Masahiro
  Nakadai, Jiabao Chen, Ken~M. Nakanishi, Kosuke Mitarai, Ryosuke Imai, Shiro
  Tamiya, Takahiro Yamamoto, Tennin Yan, Toru Kawakubo, Yuya~O. Nakagawa, Yohei
  Ibe, Youyuan Zhang, Hirotsugu Yamashita, Hikaru Yoshimura, Akihiro Hayashi,
  and Keisuke Fujii.
\newblock Qulacs: a fast and versatile quantum circuit simulator for research
  purpose.
\newblock \emph{Quantum}, 5:\penalty0 559, October 2021.
\newblock ISSN 2521-327X.
\newblock \doi{10.22331/q-2021-10-06-559}.
\newblock URL \url{http://dx.doi.org/10.22331/q-2021-10-06-559}.

\bibitem[Preskill(2018)]{Preskill_2018}
John Preskill.
\newblock Quantum computing in the nisq era and beyond.
\newblock \emph{Quantum}, 2:\penalty0 79, August 2018.
\newblock ISSN 2521-327X.
\newblock \doi{10.22331/q-2018-08-06-79}.
\newblock URL \url{http://dx.doi.org/10.22331/q-2018-08-06-79}.

\bibitem[Viamontes et~al.(2003)Viamontes, Markov, and
  Hayes]{viamontes2003improvinggatelevelsimulationquantum}
George~F. Viamontes, Igor~L. Markov, and John~P. Hayes.
\newblock Improving gate-level simulation of quantum circuits, 2003.
\newblock URL \url{https://arxiv.org/abs/quant-ph/0309060}.

\bibitem[Aaronson and Gottesman(2004)]{Aaronson_2004}
Scott Aaronson and Daniel Gottesman.
\newblock Improved simulation of stabilizer circuits.
\newblock \emph{Physical Review A}, 70\penalty0 (5), November 2004.
\newblock ISSN 1094-1622.
\newblock \doi{10.1103/physreva.70.052328}.
\newblock URL \url{http://dx.doi.org/10.1103/PhysRevA.70.052328}.

\bibitem[Calderbank et~al.(1997)Calderbank, Rains, Shor, and
  Sloane]{error_correction}
A.~R. Calderbank, E.~M. Rains, P.~W. Shor, and N.~J.~A. Sloane.
\newblock Quantum error correction and orthogonal geometry.
\newblock \emph{Physical Review Letters}, 78\penalty0 (3):\penalty0 405–408,
  January 1997.
\newblock ISSN 1079-7114.
\newblock \doi{10.1103/physrevlett.78.405}.
\newblock URL \url{http://dx.doi.org/10.1103/PhysRevLett.78.405}.

\bibitem[Gottesman(1997)]{Gottesman1997}
Daniel Gottesman.
\newblock Stabilizer codes and quantum error correction, 1997.
\newblock URL \url{https://arxiv.org/abs/quant-ph/9705052}.

\bibitem[Gottesman(1998)]{Gottesman1998}
Daniel Gottesman.
\newblock Theory of fault-tolerant quantum computation.
\newblock \emph{Physical Review A}, 57\penalty0 (1):\penalty0 127–137,
  January 1998.
\newblock ISSN 1094-1622.
\newblock \doi{10.1103/physreva.57.127}.
\newblock URL \url{http://dx.doi.org/10.1103/PhysRevA.57.127}.

\bibitem[Bravyi and Kitaev(2005)]{Bravyi2005}
Sergey Bravyi and Alexei Kitaev.
\newblock Universal quantum computation with ideal {C}lifford gates and noisy
  ancillas.
\newblock \emph{Physical Review A}, 71\penalty0 (2), feb 2005.
\newblock ISSN 1094-1622.
\newblock \doi{10.1103/physreva.71.022316}.
\newblock URL \url{http://dx.doi.org/10.1103/PhysRevA.71.022316}.

\bibitem[Knill et~al.(2008)Knill, Leibfried, Reichle, Britton, Blakestad, Jost,
  Langer, Ozeri, Seidelin, and Wineland]{Knill2008}
E.~Knill, D.~Leibfried, R.~Reichle, J.~Britton, R.~B. Blakestad, J.~D. Jost,
  C.~Langer, R.~Ozeri, S.~Seidelin, and D.~J. Wineland.
\newblock Randomized benchmarking of quantum gates.
\newblock \emph{Physical Review A}, 77\penalty0 (1), jan 2008.
\newblock ISSN 1094-1622.
\newblock \doi{10.1103/physreva.77.012307}.
\newblock URL \url{http://dx.doi.org/10.1103/PhysRevA.77.012307}.

\bibitem[Huang et~al.(2020)Huang, Kueng, and Preskill]{Huang2020}
Hsin-Yuan Huang, Richard Kueng, and John Preskill.
\newblock Predicting many properties of a quantum system from very few
  measurements.
\newblock \emph{Nature Physics}, 16\penalty0 (10):\penalty0 1050–1057, jun
  2020.
\newblock ISSN 1745-2481.
\newblock \doi{10.1038/s41567-020-0932-7}.
\newblock URL \url{http://dx.doi.org/10.1038/s41567-020-0932-7}.

\bibitem[Bravyi and Maslov(2021)]{Bravyi_2021}
Sergey Bravyi and Dmitri Maslov.
\newblock Hadamard-free circuits expose the structure of the clifford group.
\newblock \emph{IEEE Transactions on Information Theory}, 67\penalty0
  (7):\penalty0 4546–4563, July 2021.
\newblock ISSN 1557-9654.
\newblock \doi{10.1109/tit.2021.3081415}.
\newblock URL \url{http://dx.doi.org/10.1109/TIT.2021.3081415}.

\bibitem[Biamonte and Bergholm(2017)]{biamonte2017tensornetworksnutshell}
Jacob Biamonte and Ville Bergholm.
\newblock Tensor networks in a nutshell, 2017.
\newblock URL \url{https://arxiv.org/abs/1708.00006}.

\bibitem[Schollwöck(2011)]{Schollw_ck_2011}
Ulrich Schollwöck.
\newblock The density-matrix renormalization group in the age of matrix product
  states.
\newblock \emph{Annals of Physics}, 326\penalty0 (1):\penalty0 96–192,
  January 2011.
\newblock ISSN 0003-4916.
\newblock \doi{10.1016/j.aop.2010.09.012}.
\newblock URL \url{http://dx.doi.org/10.1016/j.aop.2010.09.012}.

\bibitem[Gray(2018)]{quimb}
Johnnie Gray.
\newblock \emph{quimb}: a python library for quantum information and many-body
  calculations.
\newblock \emph{Journal of Open Source Software}, 3\penalty0 (29):\penalty0
  819, 2018.
\newblock \doi{10.21105/joss.00819}.

\end{thebibliography}

\end{document}